\documentclass{article}
\usepackage{graphicx} 
\usepackage{amsmath}
\usepackage[left=3cm,right=3cm,top=3cm,bottom=3cm]{geometry}

\usepackage{pdfpages}
\usepackage{hyperref}
\usepackage{listings}
\usepackage{xcolor}
\usepackage{setspace}
\usepackage{tocloft}
\usepackage{amsmath}
\usepackage{chngcntr}
\usepackage[toc,page]{appendix}
\usepackage[T1]{fontenc}
\usepackage[nottoc]{tocbibind}
\usepackage[compact]{titlesec}
\usepackage[utf8]{inputenc}
\usepackage{natbib}
\usepackage{amsmath}
\usepackage{amssymb}
\usepackage{amsfonts}

\usepackage{multirow}
\usepackage{setspace}
\usepackage{array}
\usepackage{float}

\newcolumntype{L}[1]{>{\raggedright\let\newline\\\arraybackslash\hspace{0pt}}m{#1}}
\newcolumntype{C}[1]{>{\centering\let\newline\\\arraybackslash\hspace{0pt}}m{#1}}
\newcolumntype{R}[1]{>{\raggedleft\let\newline\\\arraybackslash\hspace{0pt}}m{#1}}

\usepackage{subfig}
\usepackage{bm}
\usepackage{multirow}
\usepackage{tikz}
\usetikzlibrary{shapes,arrows}
\usepackage{blkarray}
\usepackage[affil-it]{authblk}
\usepackage{listings}
\usepackage{siunitx}
\usepackage{lineno}
\usepackage{yhmath}
\usepackage{bbm}

\usepackage{natbib}
\usepackage{color}
\usepackage{algpseudocode}
\usepackage{algorithm2e}
\RestyleAlgo{ruled}

\providecommand{\keywords}[1]{\textbf{\textit{Keywords ---}} #1}

\title{A mixture of distributed lag non-linear models to account for spatially heterogeneous exposure-lag-response associations}

\author[1*]{Álvaro Briz-Redón}
\author[1]{Ana Corberán-Vallet}
\author[1]{Adina Iftimi}
\author[1]{Carmen Íñiguez}

\affil[1]{Department of Statistics and Operations Research, University of Valencia, Valencia, Spain}
\affil[*]{Corresponding author (\texttt{alvaro.briz@uv.es})}

\begin{document}

\maketitle

\begin{abstract}
Environmental exposures, such as air pollution and extreme temperatures, have complex effects on human health. These effects are often characterized by non-linear exposure-lag-response relationships and delayed impacts over time. Accurately capturing these dynamics is crucial for informing public health interventions. The Distributed Lag Non-Linear Model (DLNM) is a flexible statistical framework for estimating such effects in epidemiological research. However, standard DLNM implementations typically assume a homogeneous exposure-lag-response association across the study region, overlooking potential spatial heterogeneity, which can lead to biased risk estimates. To address this limitation, we introduce DLNM-Clust: a novel mixture of DLNMs that extends the traditional DLNM. Within a Bayesian framework, DLNM-Clust probabilistically assigns each geographic unit to one of $C$ latent spatial clusters, each of which is defined by a distinct DLNM specification. This approach allows capturing both common patterns and singular deviations in the exposure-lag-response surface. We demonstrate the method using municipality-level time-series data on the relationship between air pollution and the incidence of COVID-19 in Belgium. Our results emphasize the importance of spatially aware modeling strategies in environmental epidemiology, facilitating region-specific risk assessment and supporting the development of targeted public health initiatives.
\end{abstract}



\section{Introduction}

Environmental exposures, such as air pollution and extreme temperatures, exert complex effects on human health, often manifesting not only through non-linear exposure-lag-response relationships but also with delayed impacts that may span several days or even weeks. Accurately modeling these dynamics is crucial for assessing environmental health risks and informing effective public health interventions. To address this complexity, the Distributed Lag Non-Linear Model (DLNM) introduced by \cite{gasparrini2010distributed} has emerged as a robust and flexible statistical framework capable of simultaneously capturing non-linear exposure-lag-response associations and distributed lag structures over time.

DLNMs extend traditional distributed lag models by incorporating smooth functions that allow the exposure-lag-response relationship to vary with both the magnitude of exposure and the time elapsed since exposure. This dual flexibility makes DLNMs particularly well-suited for time-series analyses in environmental epidemiology, where exposures such as temperature, air pollutants, and humidity can influence health outcomes with heterogeneous latency periods and magnitudes.

A growing body of literature has demonstrated the value of DLNMs in this context. For example, \cite{Gasparrini2017} applied DLNMs to project temperature-related mortality under future climate scenarios, while \cite{he2022effects} used them to estimate the effect of night-time warming on mortality burden. Dozens of studies have employed DLNMs to examine the impact of temperature, pollution, or humidity in human health, underscoring their versatility across diverse environmental and epidemiological domains \citep[e.g.,][]{duan2019risk,guo2011impact,huang2023association,li2015particulate}.

In recent years, multi-location studies have gained importance. These studies provide information on exposure and response over a period of time for a set of multiple spatial locations, which may be cities in the same country or cities spread across the globe. The standard methodological framework in these cases is a two-stage approach, in which the exposure-lag-response association is first estimated for each spatial location, and then these estimates are combined through a meta-analysis in a second step \citep{gasparrini2012multivariate,masselot2025modelling}. This approach alleviates the computational cost that would be incurred in performing the whole analysis with a single model with multilevel effects. 

Similarly, another multi-location scenario is that corresponding to small-area studies, where both the exposure and the response are measured over a set of connected areal units. Spatial modeling serves as a cornerstone for analyzing such data, providing the analytical tools necessary to account for the geographic heterogeneity inherent in disease distribution \citep{kirby2017advances}. Over the past few decades, these methods have successfully evolved beyond simple mapping to explicitly incorporate spatial dependence, that is, the principle that health outcomes in neighboring areas are often related \citep{best2005comparison}.

Under the DLNM framework, the integration of spatial random effects has been explored as well to account for the spatial dependence in the response variable \citep{lowe2021combined}. In this line, \cite{quijal2024spatial} have introduced the Spatial Bayesian DLNM (SB-DLNM), which incorporates spatial random effects into the vector of coefficients that determine the shape of the exposure-lag-response surface. Thus, this approach enables estimating area-specific associations, which actually represent spatially-smoothed deviations from the global exposure-lag-response surface. A similar modeling framework for estimating the spatially varying nonlinear effects of heat exposure on mortality has also been recently employed by \cite{chen2025modelling}. Besides, in a related approach focusing on pregnancy outcomes, \cite{warren2020spatially} proposed the SpGPCW framework for critical window identification. This method relies on a spatially-varying Gaussian process to allow lagged health effect parameters to differ across regions, demonstrating that accounting for area-level correlation uncovers associations often hidden by models assuming spatial homogeneity.

Both the meta-analytic framework and the SB-DLNM assume that there is a single exposure-lag-response underlying association, from which each specific location is allowed to deviate moderately. In practice, this assumption implies that the exposure-lag-response relationship is rather coherent across geographic areas. However, this could be violated under certain real-world conditions. Spatial heterogeneity, arising from differences in demographics, urban form, baseline health conditions, or socioeconomic status, can generate markedly distinct local responses to the same environmental exposure. Ignoring this variability could eventually lead to biased estimates and misleading interpretations.

For this reason, we propose a novel extension of the DLNM: a Bayesian mixture model of DLNMs with an embedded clustering structure, which we call DLNM-Clust. In this framework, each area (e.g., region or municipality) is probabilistically assigned to one of $C$ latent clusters, with each cluster characterized by its own DLNM specification. This structure enables the model to capture both global exposure-lag-response associations shared across the majority of regions and local patterns unique to specific clusters of areas. Hence, while previous approaches focus on the estimation of continuous spatially varying relationships, our model provides a distinct methodological shift by discretizing these variations into latent clusters of areas with shared exposure-lag-response profiles.

The proposed modeling strategy offers several advantages. First, it enables the estimation of region-specific exposure-lag-response functions, reflecting differences in vulnerability or contextual exposures. Second, it uncovers latent spatial patterns, potentially revealing groups of regions with similar response profiles. Third, adopting a Bayesian framework provides principled inference on both parameter estimates and cluster assignments, fully accounting for uncertainty. To the best of the authors' knowledge, this is the first mixture modeling proposal in this regard in the context of the DLNM framework, even though some authors have already used mixture distributions within the DLNM framework for dealing with outliers \citep{economou2025flexible}. 

Another study related to our proposal is that of \cite{wang2023detecting}, where a novel scan statistic is defined to identify spatial clusters of locations with similar exposure-lag-response associations. In this case, the authors follow the standard methodology of scan methods \citep{kulldorff1997spatial}, which in this context is aimed at finding areas that represent the center of a cluster (of a certain radius) within which the exposure-lag-response association is different from that of the other units under study. This requires estimating such association for each of the involved areas independently and then performing likelihood ratio tests under the null hypothesis that no cluster exists for each candidate cluster. In contrast, our proposal consists of modeling the data from all areas and identifying clusters simultaneously.

The remainder of the paper is organized as follows. Section \ref{methodology} describes the standard DLNM framework for spatially-indexed data (Section \ref{methodology_1}) and introduces our proposed mixture model extension, DLNM-Clust, with two alternative clustering specifications (Section \ref{methodology_2}). Section \ref{case} presents a case study analyzing the effect of Black Carbon exposure on COVID-19 incidence in Belgium using both the classical DLNM and our proposed model. Finally, Section \ref{discussion} summarizes the main findings and discusses their implications for environmental epidemiology and public health.

\section{Methodology}
\label{methodology}

\subsection{A distributed lag non-linear model}
\label{methodology_1}

The DLNM is a statistical framework used to investigate the relationship between a time-series exposure and an outcome variable, while accounting for delayed and non-linear effects. In this section, we describe a DLNM with a rather general form that has already been used in the literature for the analysis of spatially-indexed time series. In particular, we assume a count response variable indexed in space and time, $\boldsymbol{Y}=(Y_{it})_{ \substack{ i=1, \dots,n \\ t=1, \dots,T } }$, where $i$ and $t$ denote the indices of the spatial and temporal units, respectively. This response is modeled through a negative binomial (NB) distribution, $Y_{it} \sim \text{NB}(p_{it},r)$, in order to account for the potential overdispersion of the data. We choose this distribution based on the case study that will be shown later, but the methodological framework would be equally applicable to a response variable with a Bernoulli, Gaussian, or Poisson distribution, for example.

In the NB distribution, the success probability parameter, $p_{it}$, can be related to the expected count, $E(Y_{it})=\lambda_{it}$, via the equality $p_{it} = \frac{r}{r + \lambda_{it}}$. This expression enables us to consider a reparameterization of the probability distribution in terms of $\lambda_{it}$ and $r$, the dispersion parameter, which is involved in the variance of the distribution, $Var(Y_{it})=\frac{\lambda_{it}(\lambda_{it}+r)}{r}$. 

Hence, this reparameterization allows us to specify the structure of $\lambda_{it}$ straightforwardly in terms of diverse fixed and random effects. In particular, if $\boldsymbol{x}_{it} = [x_{it}, \dots, x_{it-L}]^\top$ represents the vector of lagged exposure values observed in area $i$ at time $t$, up to a maximum lag $L$, a standard DLNM can be written as follows
\begin{equation}
    \log(\lambda_{it}) = \log(N_i) + \alpha + s(\boldsymbol{x}_{it};\boldsymbol{\eta})\text{,}
\label{model1}
\end{equation}
where $N_i$ is an offset term that denotes the population size for area $i$ (we assume constant area-level population sizes over the study period), $\alpha$ is an intercept parameter, and $s(\boldsymbol{x}_{it};\boldsymbol{\eta})$ is the function defining the exposure-lag-response association, with $\boldsymbol{\eta}$ being a vector of parameters. The complexity of $s$ (and therefore the number of parameters contained in $\boldsymbol{\eta}$) depends on our choice of basis functions for both the exposure and lag dimensions. In particular, if the dimensions of the basis functions are $v_x$ and $v_{\ell}$, respectively, a cross-basis of dimension $v_x \cdot v_{\ell}$ is generated, allowing us to specify $s(\boldsymbol{x}_{it}; \boldsymbol{\eta}) = \boldsymbol{b}_{it}^\top \boldsymbol{\eta}$, where $\boldsymbol{b}_{it}$ denotes the cross-basis vector obtained by applying the corresponding basis functions to the lagged exposure vector, $\boldsymbol{x}_{it}$, at time $t$. Thus, the term $s(\boldsymbol{x}_{it}; \boldsymbol{\eta})$ encapsulates the effect of the lagged exposure values observed in area $i$ at time $t$ on the response. Further details on the algebraic specification of the DLNM through cross-basis functions can be found in the main references about this model \citep{gasparrini2010distributed,gasparrini2014modeling}.



To account for the spatial and temporal variability in the outcome variable that might be unexplained by the specified exposure-lag-response association, we can extend the DLNM in Equation \ref{model1} by incorporating spatial and temporal random effects \citep{lowe2021combined,rutten2025lagged}. Specifically, we can add structured and unstructured area-level spatial random effects, denoted by $u_i$ and $v_i$, respectively, following the Besag-York-Molliè (BYM) model \citep{besag1991bayesian}, and a structured temporal random effect, $\gamma_t$
\begin{equation}
\log(\lambda_{it}) = \log(N_i) + \alpha + s(\boldsymbol{x}_{it};\boldsymbol{\eta}) + u_i + v_i + \gamma_{t} \text{.}
\label{model2}
\end{equation}

First, the structured spatial random effects $u_i$ follow a Conditional Autoregressive (CAR) specification
$$u_i | u_{-i} \sim \text{Normal}\left(\sum_{j\neq i} w_{ij} u_j, \frac{\sigma_u^2}{n(i)}\right)\text{,}$$
where $n(i)$ is the number of neighbors for area $i$, $w_{ij}$ is the $(i,j)$ element of the row-normalized neighborhood matrix, and $\sigma^2_{u}$ represents the variance of this random effect. Here, we will assume a contiguity-based row-normalized neighborhood matrix, meaning that areas $i$ and $j$ are neighbors if they share some geographical boundary \citep{briz2022comparison}. Second, the unstructured spatial random effects $v_i$ are assumed to be independent and identically distributed, $v_i \sim \text{Normal}(0, \sigma_v^2)$, where $\sigma_v^2$ is the variance of the unstructured random effect. We note that while the individual terms in the BYM model can face identifiability issues, our primary focus in this context is to estimate the total spatial effect ($u_i + v_i$). Alternative reparameterizations of the spatial random effects, such as the BYM2 model \citep{riebler2016intuitive}, could also be considered to enhance interpretability.

Finally, the temporal random effects are specified through a second-order random walk,
$\gamma_{t}|\gamma_{t-1},\gamma_{t-2} \sim \text{Normal}(2\gamma_{t-1}-\gamma_{t-2},\sigma^2_{\gamma})$, with $\sigma^2_{\gamma}$ the variance component. We have chosen a second-order random walk to capture the temporal dependence of the data in order to examine the results of the proposed model on the data analyzed by \cite{rutten2025lagged}, who also considered this kind of effect. Instead, a first-order random walk (which induces less smoothing) or a family of natural cubic splines could also have been used to capture the temporal variability of the response.

We estimate the DLNM indicated in Equation \ref{model2} through a Bayesian inferential framework, which requires assigning prior distributions to each of the parameters involved. Following \cite{ntzoufras2011bayesian}, we choose a Gamma prior, $r\sim\text{Ga}(0.001,0.001)$ for the dispersion parameter of the NB distribution. Regarding the intercept, $\alpha$, and the parameters contained in the $\boldsymbol{\eta}$ vector, we choose $\alpha\sim \text{Normal}(0,100)$ and $\boldsymbol{\eta}\sim \text{Normal}(0,\Sigma)$ with $\Sigma=100\text{I}$ ($\text{I}$ is the identity matrix of the required size), assuming a zero-mean with large prior variance for each of the elements of $\boldsymbol{\eta}$ and independence across components. 

Besides, we consider uniform vaguely informative uniform priors on the standard deviations, $\sigma_u\sim \text{Unif}(0,L_{\sigma_{u}})$, $\sigma_v\sim \text{Unif}(0,L_{\sigma_{v}})$, and $\sigma_\gamma\sim \text{Unif}(0,L_{\sigma_{\gamma}})$, where $L_{\sigma_{u}}$, $L_{\sigma_{v}}$, and $L_{\sigma_{\gamma}}$ are upper bounds of the corresponding standard deviation, which need to be chosen based on the application and knowledge about the data. For the analysis presented later, we will consider $L_{\sigma_u} = L_{\sigma_v} = L_{\sigma_\gamma} = 10$.

\subsection{A mixture of distributed lag non-linear models}
\label{methodology_2}

In this section, we formulate a mixture of DLNMs, where each component of the mixture represents a specific DLNM, allowing so several distinct exposure-lag-response associations within the study window. We name this proposal as the DLNM-Clust model. As a first approach (Section \ref{Mixt}), we consider that the assignment of areas to clusters has no spatial dependence, whereas in Section \ref{SPMixt} we show how to induce a spatially-smoothed clustering structure. 

\subsubsection{Non-spatial cluster assignment} \label{Mixt}

Let $C$ denote the number of components in the mixture, which will be referred to from now on as clusters. Now, the mean $\lambda_{it}$ is assumed to depend on the cluster to which each areal unit $i$ belongs. Let $z_i$ be the variable that indicates the cluster assigned to unit $i$. Thus, we denote the mean of $Y_{it}$ conditional on $z_i=c$, for some $c\in\{1, \dots,C\}$, as $\lambda_{it}(c)$. A key output of the model is then the inferred cluster assignment for each of the units under study.

Therefore, we need to specify $\lambda_{it}(c)$ uniquely for $c=1, \dots,C$. In this regard, we assume a single dispersion parameter, $r$, for all clusters and shared spatial and temporal random effects. Thus, the specifications of the different clusters only differ in terms of the coefficients of the $\boldsymbol{\eta}$ vectors. These assumptions lead to the following mixture of DLNMs
\begin{equation}
\begin{gathered}
Y_{it}|z_i=c \sim \text{NB}(\lambda_{it}(c),r)\\
 \log(\lambda_{it}(c)) = \log(N_i) + \alpha + s(\boldsymbol{x}_{it};\boldsymbol{\eta}_c)+ u_i + v_i + \gamma_{t}\\
z_i\sim \text{Cat}(q_{i1}, \dots,q_{iC})\\
(q_{i1}, \dots,q_{iC})\sim \text{Dir}(1, \dots,1) \text{,}
\label{eq:mixt_model}
\end{gathered}
\end{equation}
where $\boldsymbol{\eta}_{c}$ is the vector of coefficients defining the exposure-lag-response association determined by cluster $c$, and $q_{ic}$ is the probability that unit $i$ belongs to cluster $c$, $c=1, \dots,C$. Specifically, we assume that $z_i$ is a random variable following a Categorical distribution over $\{1,\dots,C\}$, where $q_{ic} = P(z_i = c)$ and $\sum_{c=1}^C q_{ic} = 1$. We further assume that all clusters are equally likely a priori, so that $P(z_i = c) = \frac{1}{C}$, for $c=1, \dots,C$, by assigning an uninformative Dirichlet prior, $\text{Dir}(1, \dots,1)$, on the vector of $q_{ic}$'s. 

Finally, for each $\boldsymbol{\eta}_c$ we consider the prior $\boldsymbol{\eta}_c\sim \text{Normal}(0,\Sigma)$, with $\Sigma=100\text{I}$, whereas for the remaining parameters, which are in common with the standard DLNM summarized in Equation \ref{model2}, we assume the same priors previously discussed.


\subsubsection{Spatially-smoothed cluster assignment}
\label{SPMixt}

Now, we propose the same type of mixture model, though inducing a spatial structure to the assignment of areas to clusters. This strategy aims to encourage the identified clusters to show greater consistency and robustness throughout the spatial study window. Moreover, this kind of clustering should facilitate the subsequent interpretation of the results and the consequent public health decision-making.

The difference from the previously proposed mixture model lies in the fact that a prior distribution with a spatial structure is assigned to the probabilities $q_{ic}$. Specifically, the BYM model used to establish spatial dependence between the observed counts can be used for the vector $(q_{i1}, \dots,q_{iC})$. We note that this implies defining a spatial structure for each cluster, for $c=1, \dots,C$. Thus, in line with multinomial models \citep{nibbering2024high,papastamoulis2023model}, we propose the following modeling framework
\begin{equation}
\begin{gathered}
z_i\sim \text{Cat}(q_{i1}, \dots,q_{iC})\\
q_{ic}=\frac{\varphi_{ic}}{\sum_{c=1}^C\varphi_{ic}}\\
\log(\varphi_{ic})=u^c_i+v^c_i\\
u^c_i | u^c_{-i} \sim \text{Normal}\left(\sum_{j\neq i} w_{ij} u^c_j, \frac{\sigma^2_{uc}}{n(i)}\right)\\
v^c_i \sim \text{Normal}\left(0, \sigma^2_{vc}\right)\text{,}
\label{eq:mixt_model_sp}
\end{gathered}
\end{equation}
where $u_i^c$ and $v_i^c$ are the spatial random effects that enable us to control the spatial dependence in the assignment of area $i$ to cluster $c$, and $\varphi_{ic}$ is an auxiliary (positive) variable that allows defining $q_{ic}$ in such a way that $\sum_{c=1}^C q_{ic}=1$, as required. Thus, by introducing a spatially smoothed structure in the definition of the $q_{ic}$ probabilities, we encourage neighboring areas to be assigned to the same cluster. The spatial dependence across the $u_i^c$ values ensures that the $\varphi_{ic}$ terms are similar for adjacent areas, which ultimately yields closer $q_{ic}$ probabilities across those locations for any given cluster $c$.

The standard deviation parameters, $\sigma_{uc}$ and $\sigma_{vc}$, for $c=1, \dots,C$, are assigned HalfNormal(0,1) priors. Hence, in this case, a prior with a lower range of variation for the standard deviations is chosen to avoid potential identifiability issues. The prior distribution of each of the other parameters involved in Equation \ref{eq:mixt_model} is chosen analogously, leading us to the following full specification of the proposed DLNM-Clust model with spatially-smoothed cluster assignment
\begin{equation}
\begin{gathered}
Y_{it}|z_i=c \sim \text{NB}(\lambda_{it}(c),r)\\
 \log(\lambda_{it}(c)) = \log(N_i) + \alpha + s(\boldsymbol{x}_{it};\boldsymbol{\eta}_c)+ u_i + v_i + \gamma_{t}\\
 r\sim\text{Ga}(0.001,0.001)\\
 \alpha\sim \text{Normal}(0,100)\\
 \boldsymbol{\eta}_c\sim \text{Normal}(0,100\text{I})\\
 u_i | u_{-i} \sim \text{Normal}\left(\sum_{j\neq i} w_{ij} u_j, \frac{\sigma_u^2}{n(i)}\right), \sigma_{u}\sim\text{Unif}(0,10)\\
 v_i \sim \text{Normal}(0, \sigma_v^2), \sigma_{v}\sim\text{Unif}(0,10)\\
 \gamma_{t}|\gamma_{t-1},\gamma_{t-2} \sim \text{Normal}(2\gamma_{t-1}-\gamma_{t-2},\sigma^2_{\gamma}), \sigma_{\gamma}\sim\text{Unif}(0,10)\\
z_i\sim \text{Cat}(q_{i1}, \dots,q_{iC})\\
q_{ic}=\frac{\varphi_{ic}}{\sum_{c=1}^C\varphi_{ic}}\\
\log(\varphi_{ic})=u^c_i+v^c_i\\
u^c_i | u^c_{-i} \sim \text{Normal}\left(\sum_{j\neq i} w_{ij} u^c_j, \frac{\sigma^2_{uc}}{n(i)}\right), \sigma_{uc}\sim \text{HalfNormal}(0,1)\\
v^c_i \sim \text{Normal}\left(0, \sigma^2_{vc}\right),\sigma_{vc}\sim \text{HalfNormal}(0,1) \text{.}
\label{eq:full_model}
\end{gathered}
\end{equation}

\subsection{Model comparison and selection of the number of clusters}
\label{model_selection}

Under a Bayesian framework, model assessment and comparison can be conducted through some well-known metrics for measuring both goodness-of-fit and model complexity, such as the Deviance Information Criterion (DIC) of \cite{spiegelhalter2002bayesian}, or the Watanabe-Akaike Information Criterion (WAIC) proposed by \cite{watanabe2010asymptotic}. In our case, the WAIC, which accounts for the effective number of parameters to control for model overfitting, has been used for the comparison. In particular, we have employed the specific version of WAIC (pWAIC2) proposed by \cite{gelman2014understanding}, which is recommended for being closer in practice to a leave-one-out cross-validation analysis. In short, a reduction in the WAIC indicates superior model performance.



The WAIC allows for model comparison in a global sense. For conducting a local analysis of model suitability, \cite{hoayek2025assessing} have recently studied the use of Shannon's entropy from information theory for quantifying how well a clustering procedure assigns each observation to a specific cluster. Specifically, for a given observation, $i$, the entropy is computed as
\begin{equation}
H(i)=-\sum_{c=1}^{C} p_c(i)\log_2(p_c(i))\text{,}
\end{equation}
where $p_c(i)$ is the probabilistic degree of membership of observation $i$ to cluster $c$. By convention, if $p_c(i)=0$, then $p_c(i)\log_2(p_c(i))=0$. This metric directly quantifies the level of uncertainty in the clustering outcome. Specifically, an entropy value approaching its theoretical minimum of zero indicates that the observation is assigned with high certainty to one of the clusters. Conversely, higher entropy values suggest that the observation is equally likely to belong to multiple clusters, which reflects the existence of large uncertainty about the clustering process. The theoretical maximum value of $H(i)$ corresponds to $-\sum_{c=1}^{C} \frac{1}{C}\log_2(\frac{1}{C})$, where we assume $p_c(i)=\frac{1}{C}$, $c=1, \dots,C$, which represents that the observation is assigned to all of the clusters with equal probability. Therefore, the objective is to find a DLNM-Clust model, for some value of $C$, that favors small entropies, suggesting that the partitioning of the study area into that specific number of clusters is considerably consistent and well defined.

\subsection{Model implementation and software}

The models previously described in the paper have been implemented with the NIMBLE software for Bayesian inference \citep{de2017programming}, which is based on Markov chain Monte Carlo (MCMC) routines. The 4.3.1 version of the R programming language \citep{teamR} has been used for the analysis. In particular, the R packages \textsf{dlnm} \citep{DLNM}, \textsf{ggplot2} \citep{ggplot2}, and \textsf{nimble} \citep{de2017programming} have been employed. 

\section{Case study}
\label{case}

\subsection{Data}

The COVID-19 dataset for Belgium employed in this study is based on the data recently analyzed by \cite{rutten2025lagged}. These data are suitable for the application of the DLNM-Clust model due to their high spatio-temporal granularity and the inherent socio-demographic diversity of the study area. Furthermore, the dataset serves as an ideal benchmark, as it was previously analyzed using a standard, spatially homogeneous DLNM \citep{rutten2025lagged}. This allows for a direct comparison between a global model and our proposed clustering framework, highlighting the added value of identifying localized exposure-lag-response associations that may be obscured in aggregate analyses.

The dataset can be downloaded from \href{https://github.com/Rutten-Sara/Bayesian-DLNM-Air-pollution-and-COVID-19-in-Belgium}{https://github.com/Rutten-Sara/Bayesian-DLNM-Air-pollution-and-COVID-19-in-Belgium}
. This dataset provides weekly information on newly reported COVID-19 cases and deaths across Belgian municipalities. In our analysis, we consider the number of weekly new cases as the response variable, denoted by $Y_{it}$, for $i=1,\ldots,581$ municipalities and $t=1,\ldots,71$ weeks.

In addition, the dataset includes several covariates related to air pollutants, enabling the study of associations between weekly averaged pollutant levels and COVID-19 incidence and mortality rates. Specifically, we focus on the covariate representing the concentration of Black Carbon (BC), measured in $\mu \text{g}/\text{m}^3$, and estimate its effect on COVID-19 incidence. This choice is motivated by the findings of \cite{rutten2025lagged}, who showed that models including BC achieved the best performance compared to those considering $\text{NO}_2$, $\text{O}_3$, or $\text{PM}$ levels. Nevertheless, this selection is of secondary importance, since the case study is mainly intended to illustrate the application and potential advantages of the proposed methodology. Indeed, it is worth noting that \cite{rutten2025lagged} also considered vaccination coverage as an additional covariate, modeling it with a nonlinear and distributed lag effect. For simplicity, however, we do not include this variable in the present analysis.

\subsection{Standard model (DLNM)}

First, we fitted the standard DLNM to the COVID-19 case counts in Belgian municipalities. Following \cite{rutten2025lagged}, we considered lagged effects up to 8 weeks for BC levels, natural splines with two equally spaced knots for the exposure dimension (3 parameters), and one knot plus an intercept for the lag dimension (3 parameters). This specification results in a $\boldsymbol{\eta}$ vector of length 9. The inference is conducted through a single MCMC chain of length 80000 with a burn-in period of length 40000 and a thinning of 10.

\begin{figure}
    \centering
    \includegraphics[width=0.5\linewidth,angle=-90]{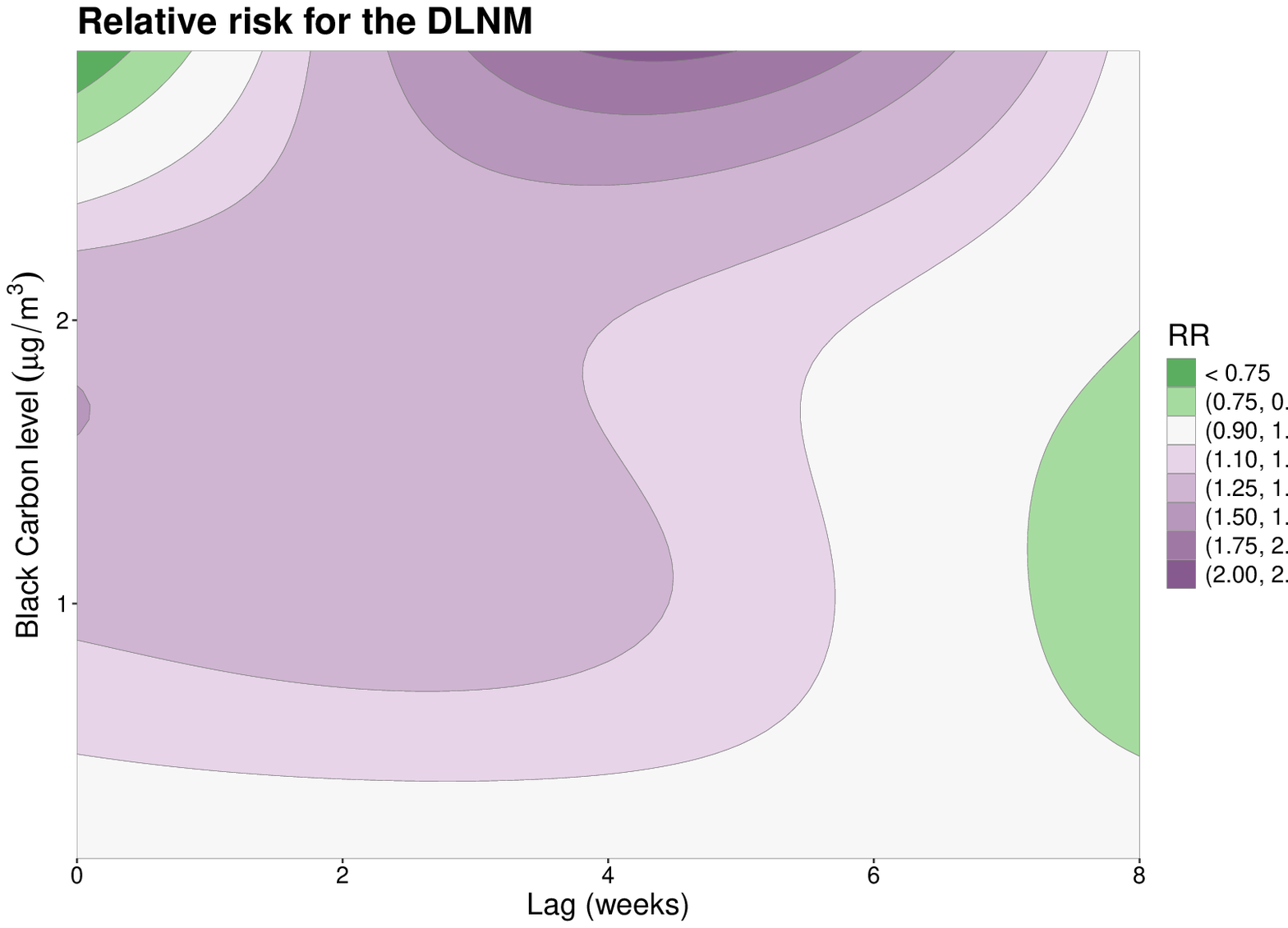}
    \caption{Contour plot of the relative risk of COVID-19 incidence as a function of BC levels at different lags (in weeks) estimated through the DLNM. Here we are considering the 5\% sample percentile as the reference value for the BC level}
    \label{fig:contour}
\end{figure}

Figure \ref{fig:contour} presents the contour plot corresponding to the estimated exposure-lag-response association. The relative risk surface was obtained using the \texttt{crosspred} function from the \texttt{dlnm} package. Specifically, the posterior mean of $\boldsymbol{\eta}$ was supplied as the \texttt{coef} parameter, while the posterior variance-covariance matrix of the parameters within $\boldsymbol{\eta}$ was provided as the \texttt{vcov} parameter. The figure shows that the highest relative risk values concentrate at high BC levels, with a temporal lag of 3–5 weeks. A smaller increase in risk is also observed for moderate pollutant levels at short lags (1 week).

\begin{figure}
    \centering
    \subfloat[]{\includegraphics[width=6.9 cm,angle=0]{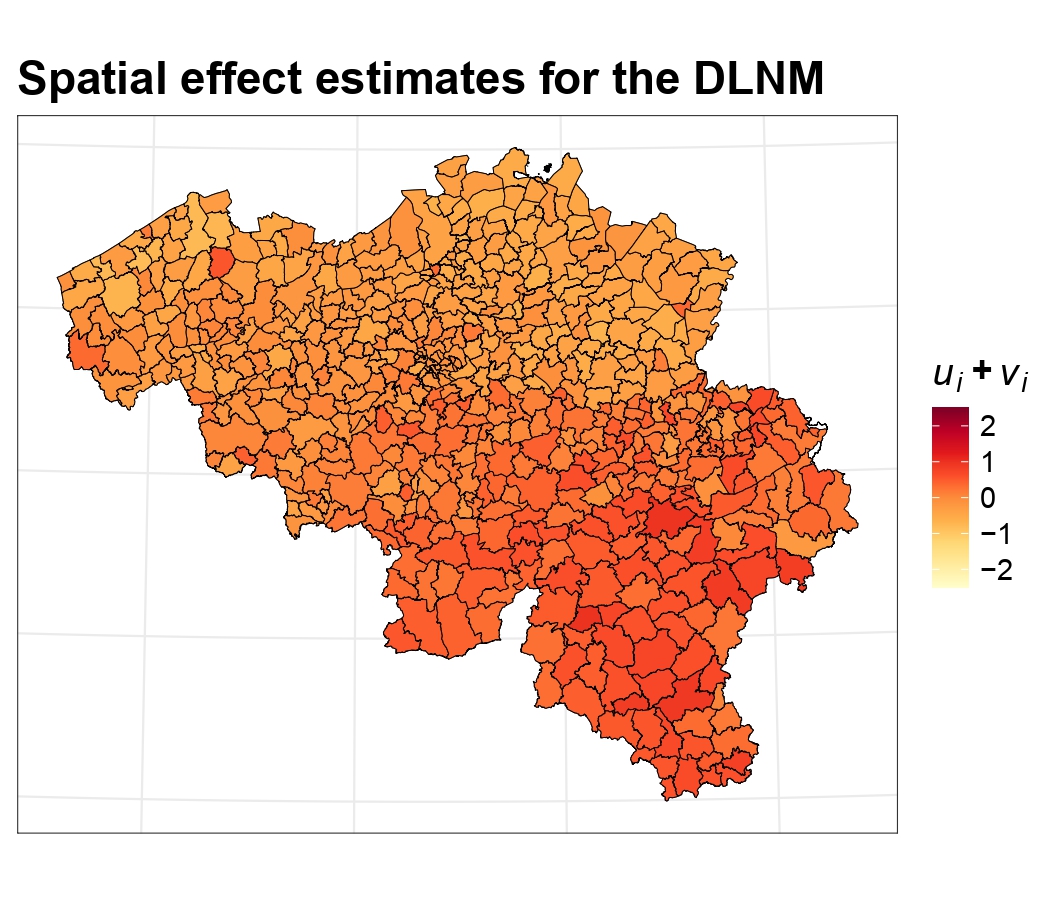}\label{fig:spatiotemporal_a}}
    \subfloat[]{\includegraphics[width=6.5cm,angle=0]{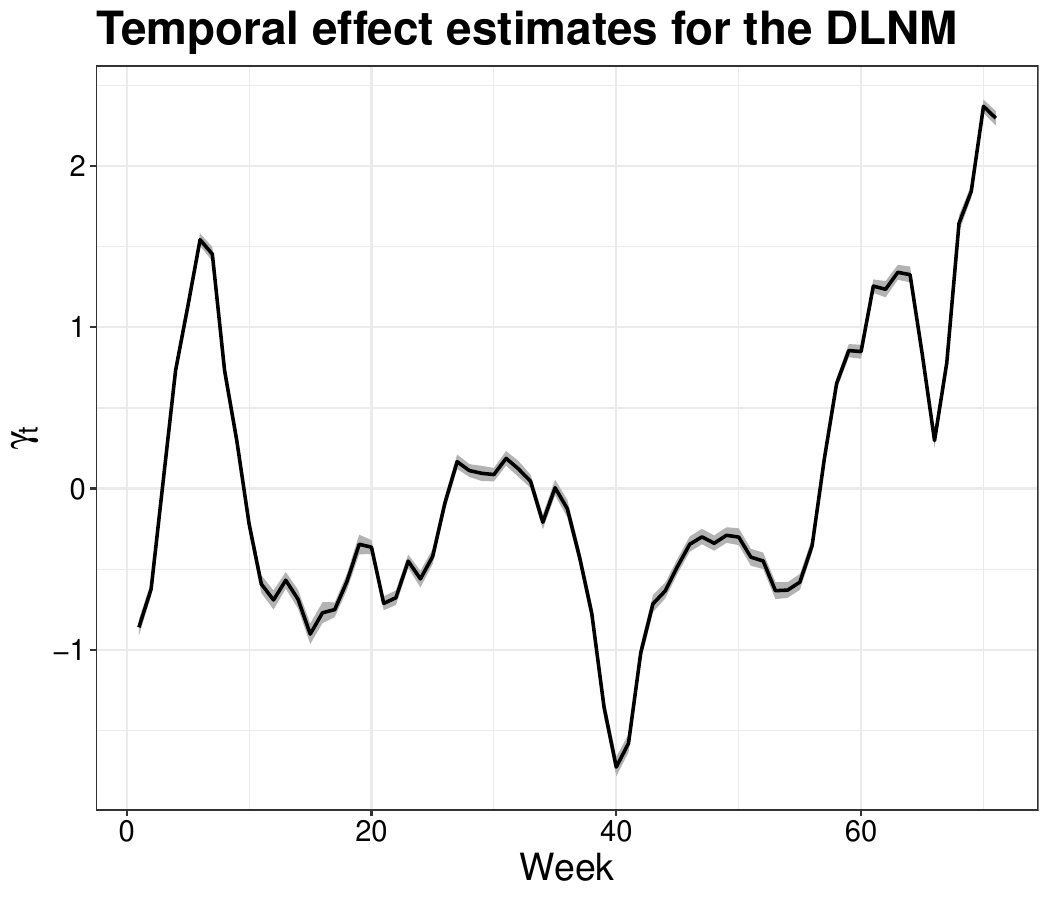}\label{fig:spatiotemporal_b}}
    \caption{Spatial effect estimates (posterior mean of $u_i+v_i$) for the 581 Belgian municipalities (a), and temporal week-level effect estimates (posterior mean of $\gamma_t$) for the 71 weeks under study (b). In (b), the 95\% credible band associated with the estimates is also provided}
    \label{fig:spatiotemporal}
\end{figure}

Figure \ref{fig:spatiotemporal} summarizes the estimated spatial and temporal effects. In Figure \ref{fig:spatiotemporal_a}, a spatial gradient can be observed towards the southeast, indicating higher infection levels in those municipalities, unrelated to pollutant exposure. This suggests the presence of unobserved covariates not accounted for in the model. Regarding temporal random effects, Figure \ref{fig:spatiotemporal_b} displays the estimated $\gamma_t$ terms with their associated credible bands. This plot captures the temporal dynamics of incidence levels during the study period, highlighting distinct peaks at specific points in time. Overall, the analysis closely replicates the findings reported by \cite{rutten2025lagged}.

\subsection{Mixture model (DLNM-Clust)}

\subsubsection{Model comparison and selection}

\begin{figure}
    \centering
    \includegraphics[width=0.5\linewidth,angle=-90]{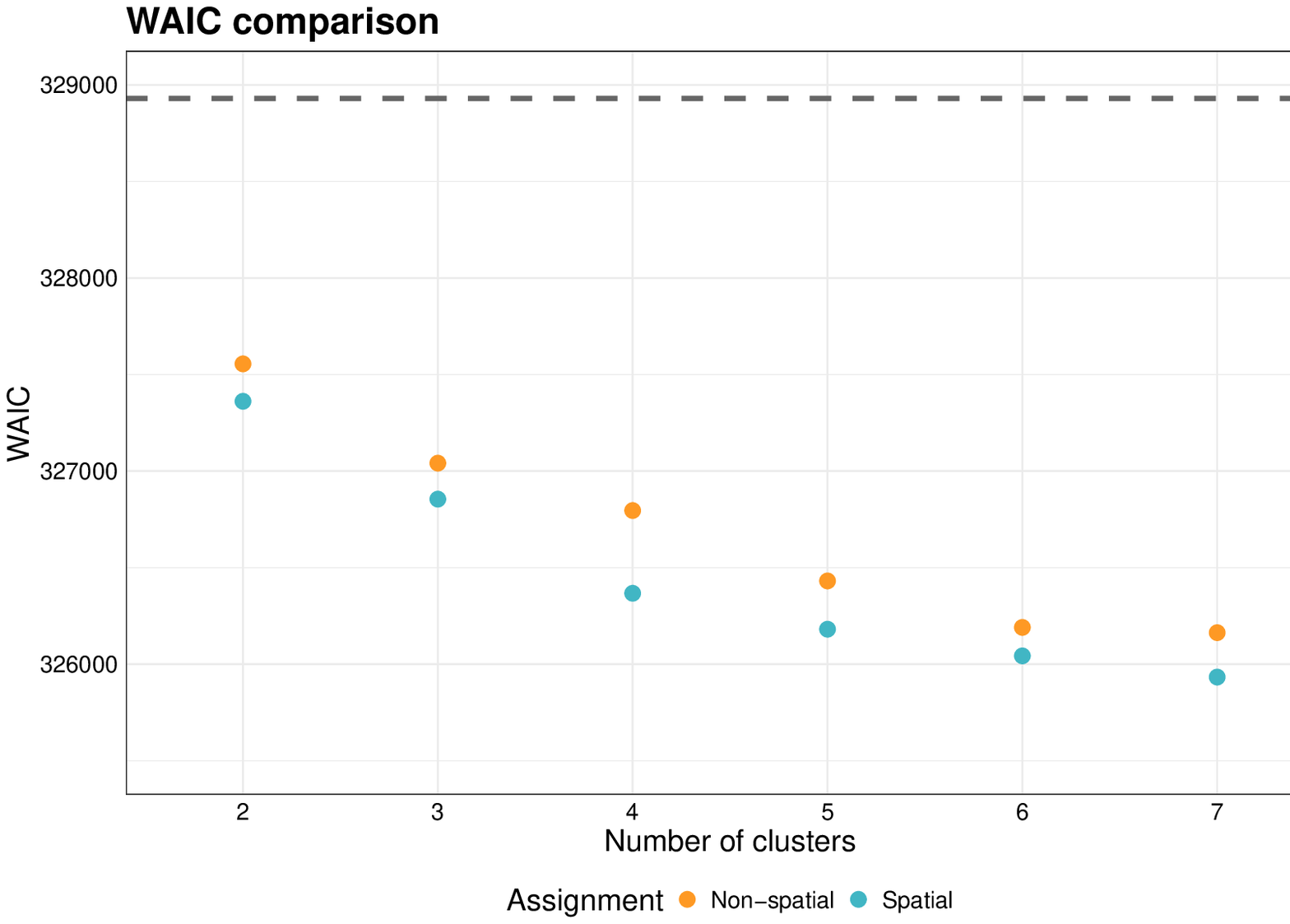}
    \caption{WAIC values presented by the DLNM-Clust models, for $C$ ranging from 2 to 7, considering both the non-spatial and the spatial assignment of areas to clusters. The dashed line represents the WAIC value of the standard DLNM}
    \label{fig:WAIC}
\end{figure}

Next, we fitted the DLNM-Clust model with values of $C$ ranging from 2 to 7. Again, we used a single MCMC chain of length 80000 with a burn-in period of length 40000 and a thinning of 10, which leads to a computational cost of about 8 hours on a personal laptop with an i7 processor. A comparative analysis was conducted between these models using the metrics described in Section \ref{model_selection}, and their performance was also contrasted with that of the standard DLNM. The WAIC results indicate that the DLNM-Clust model provides a substantial improvement over the standard model, as illustrated in Figure \ref{fig:WAIC}, where the WAIC value of the standard DLNM is shown as a dashed line. Both non-spatial and spatial clustering assignments enhance model performance; however, models incorporating spatial clustering consistently outperform their non-spatial counterparts. Moreover, the improvement in WAIC becomes less pronounced as the value of $C$ increases.

\begin{figure}
    \centering
    \includegraphics[width=0.5\linewidth,angle=-90]{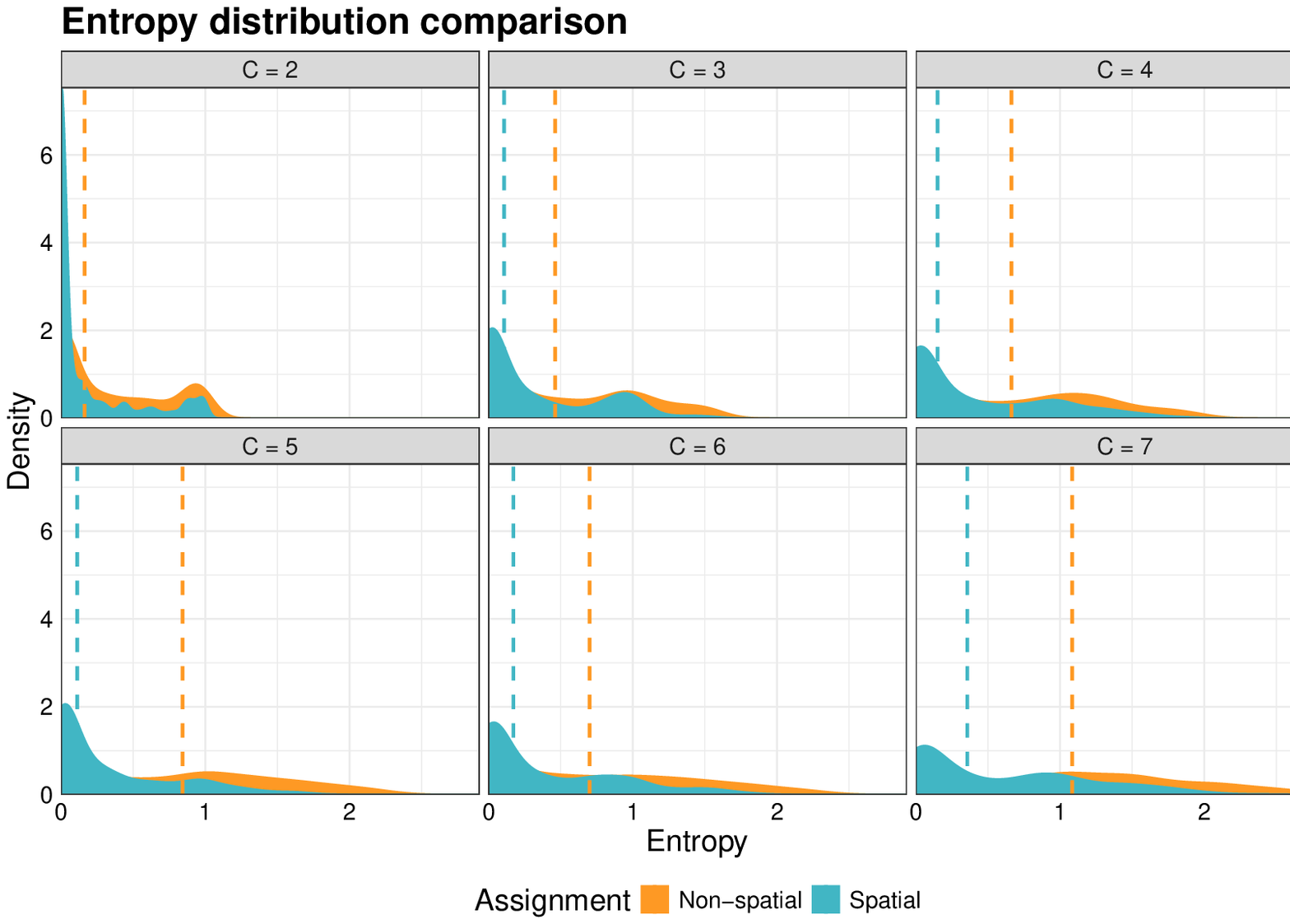}
    \caption{Density plot of entropy values obtained from the posterior probabilities $P(z_i=c|\boldsymbol{Y})$, $c=1, \dots,C$, for $C$ ranging from 2 to 7, considering both the non-spatial and the spatial assignment of areas to clusters. The dashed lines represent the median entropy value for each model type under each choice of $C$}
    \label{fig:entropy}
\end{figure}

Figure \ref{fig:entropy} presents a histogram of the entropy values for the set of observations, considering the mixture model with $C$ values ranging from 2 to 7. The posterior probabilities, $P(z_i=c|\boldsymbol{Y})$, for $c=1,\ldots,C$, were computed as the proportion of MCMC iterations in which each region was assigned to a given cluster. This approach allows us to quantify the uncertainty associated with the cluster assignments, giving us a better understanding of the relationships between municipalities and clusters.

In general, it is observed that as $C$ increases, the distribution gradually shifts to the right, indicating a larger number of observations with high entropy values. This behavior is expected, since increasing $C$ makes the model more complex, which in turn can make cluster assignment more uncertain. Conversely, for $C=2$, the mean entropy reaches its lowest value, reflecting greater certainty in the cluster assignments. Based on these results, we chose the model with $C=5$ and spatial clustering assignment as the optimal one, considering the balance between WAIC improvement and entropy levels.

\subsubsection{Main results for the selected model}

\begin{figure}
    \centering
\subfloat[]{\includegraphics[width=7.2cm,angle=0]{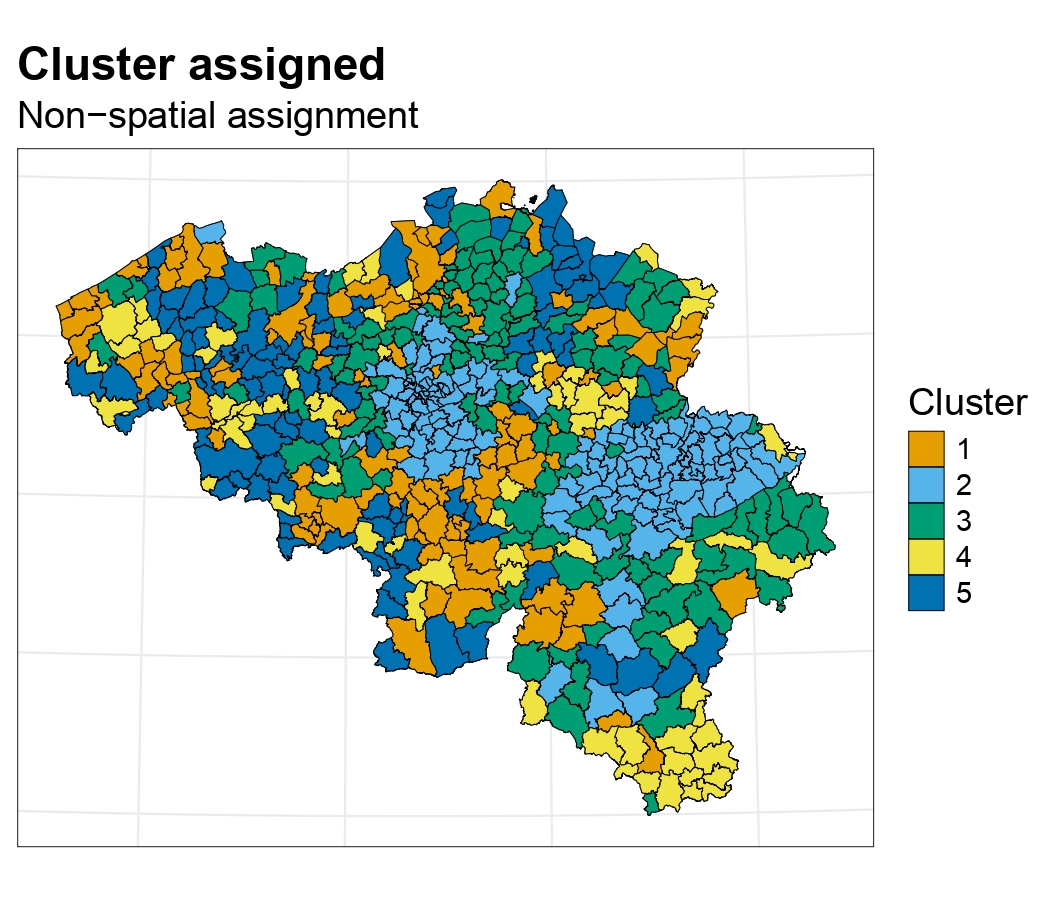}\label{fig:cluster_map_a}}
\subfloat[]{\includegraphics[width=7.2cm,angle=0]{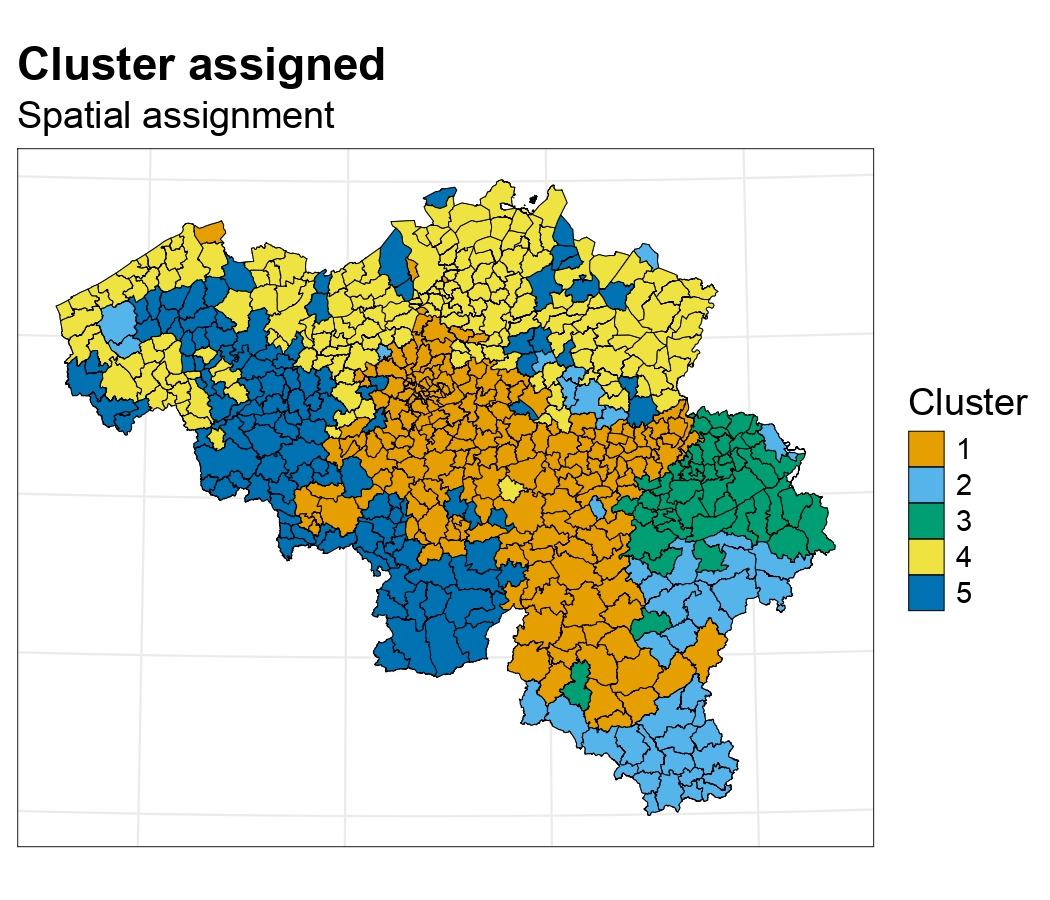}\label{fig:cluster_map_b}}\\
\subfloat[]{\includegraphics[width=7.2cm,angle=0]{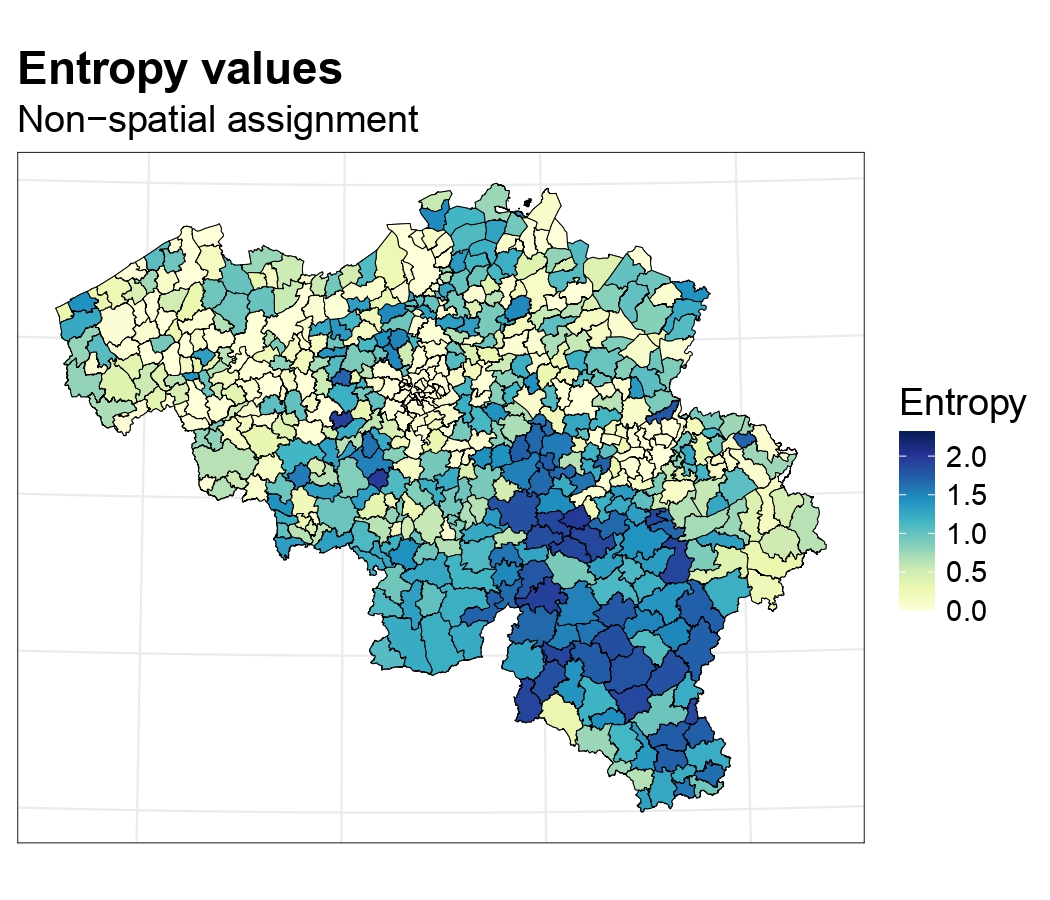}\label{fig:cluster_map_c}}
\subfloat[]{\includegraphics[width=7.2cm,angle=0]{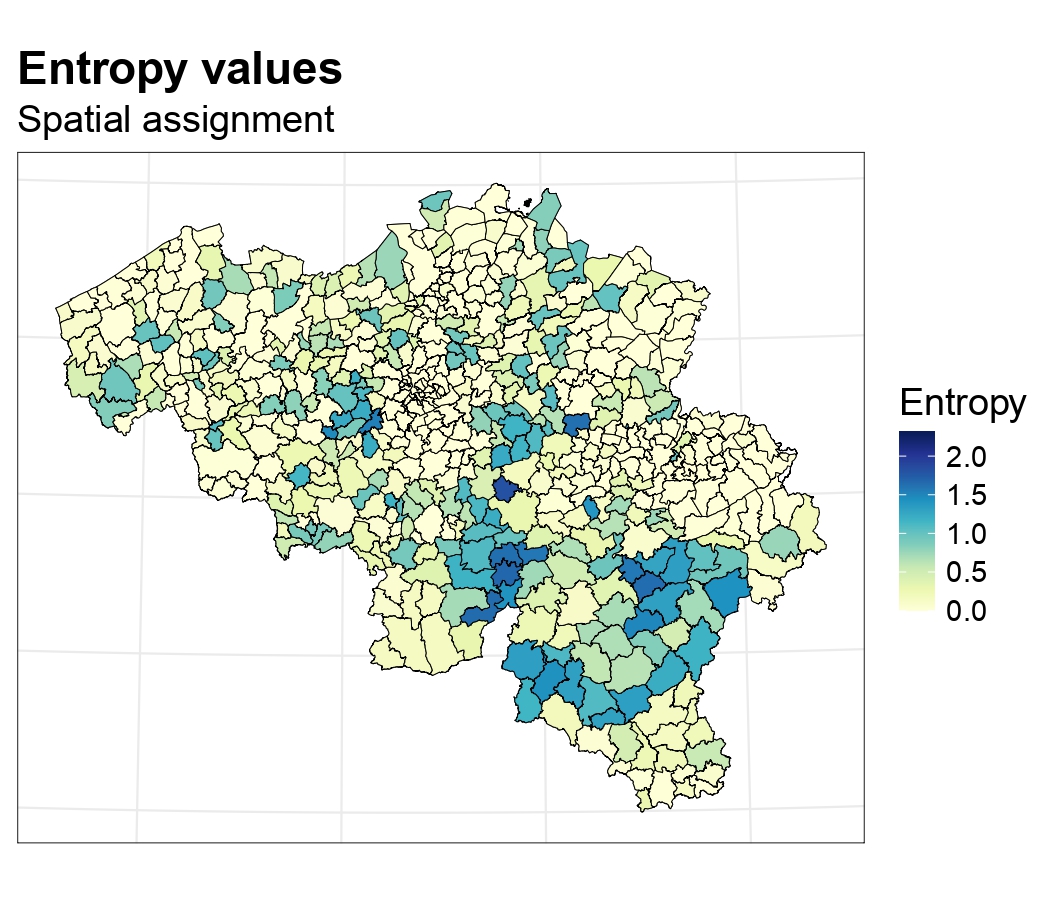}\label{fig:cluster_map_d}}\\
    \caption{Map of assigned clusters provided by the DLNM-Clust model ($C=5$), obtained as $\text{argmax}_{c=1, \dots,C} \hspace{0.1cm} P(z_i=c|\boldsymbol{Y})$, considering the non-spatial (a) and the spatial (b) assignment of areas to clusters. We also show the map of entropy values obtained from the posterior probabilities $P(z_i=c|\boldsymbol{Y})$ for the non-spatial (c) and the spatial (d) versions of this model. For the choice of $C=5$ the entropy values range from 0 to $-\sum_{c=1}^{5} \frac{1}{5}\log_2(\frac{1}{5})=2.321928$}
    \label{fig:cluster_map}
\end{figure}

In this section, we mainly focus on the results provided by the DLNM-Clust model for $C=5$ with spatial assignment of areas to clusters. However, Figure \ref{fig:cluster_map} presents the cluster assignment map considering both the non-spatial and spatial assignment versions of the model. This allows us to assess how the inclusion of a spatial structure in the clustering produces more interpretable results. Indeed, while Figure \ref{fig:cluster_map_a} does not reveal clear spatial patterns and neighboring areas are often assigned to different clusters, Figure \ref{fig:cluster_map_b} provides a more coherent representation of the spatial distribution of the clusters within the study area. In the latter case, 193, 42, 45, 179 and 122 municipalities are assigned to Clusters 1 to 5 respectively, which gives an idea of how common each cluster-specific association could be. It is also worth mentioning that the five clusters identified in Figures \ref{fig:cluster_map_a} and \ref{fig:cluster_map_b} are not directly comparable, as the estimated exposure-lag-response associations depend on the specific model fitted in each case. Occasionally, clusters may exhibit similar exposure-lag-response associations, but this is not guaranteed. 

Moreover, Figures \ref{fig:cluster_map_c} and \ref{fig:cluster_map_d} show the entropy value maps for the models with non-spatial and spatial assignment, respectively. In both cases, it can be seen that the entropy values tend to be higher in the southern part of the study window, suggesting that the municipalities in this part of Belgium are characterized with less certainty by the DLNM-Clust model. However, it is also evident that the model with spatial assignment gives rise to substantially lower entropy values, as previously indicated in Figure \ref{fig:entropy}.

\begin{figure}[htbp]
\centering
\subfloat[]{\includegraphics[width=5.2cm,angle=-90]{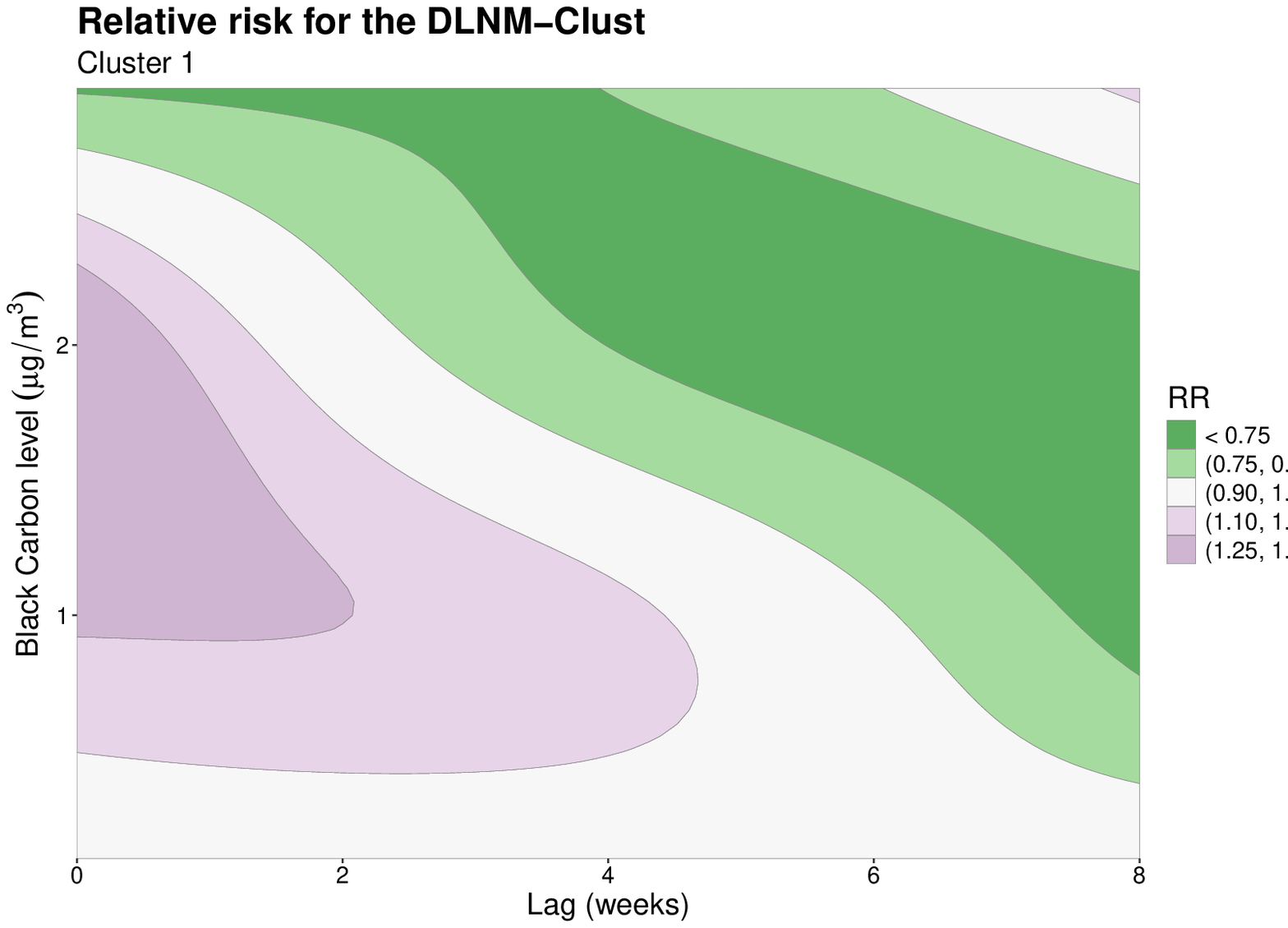}}
\subfloat[]{\includegraphics[width=5.2cm,angle=-90]{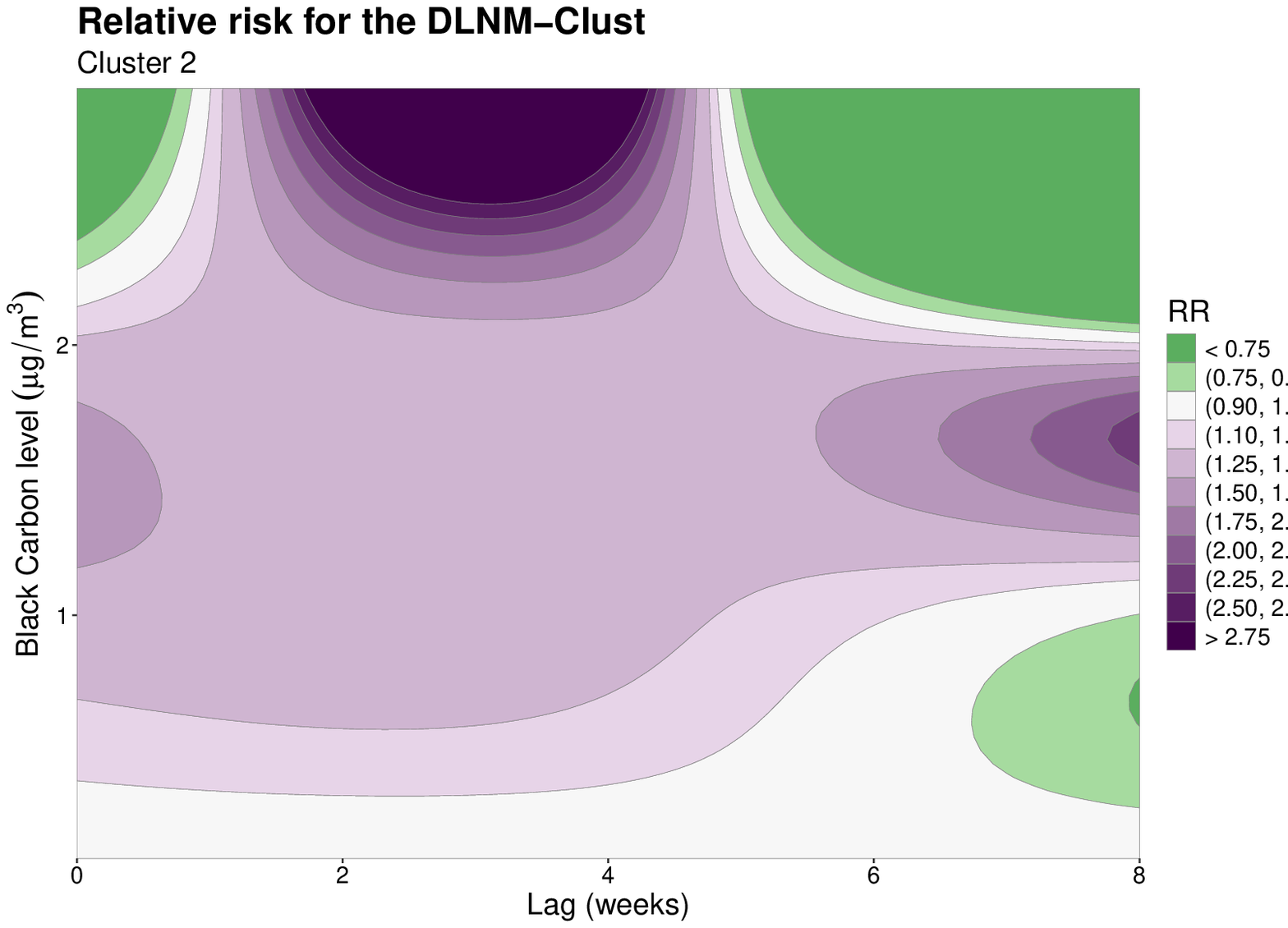}}\\
\subfloat[]{\includegraphics[width=5.2cm,angle=-90]{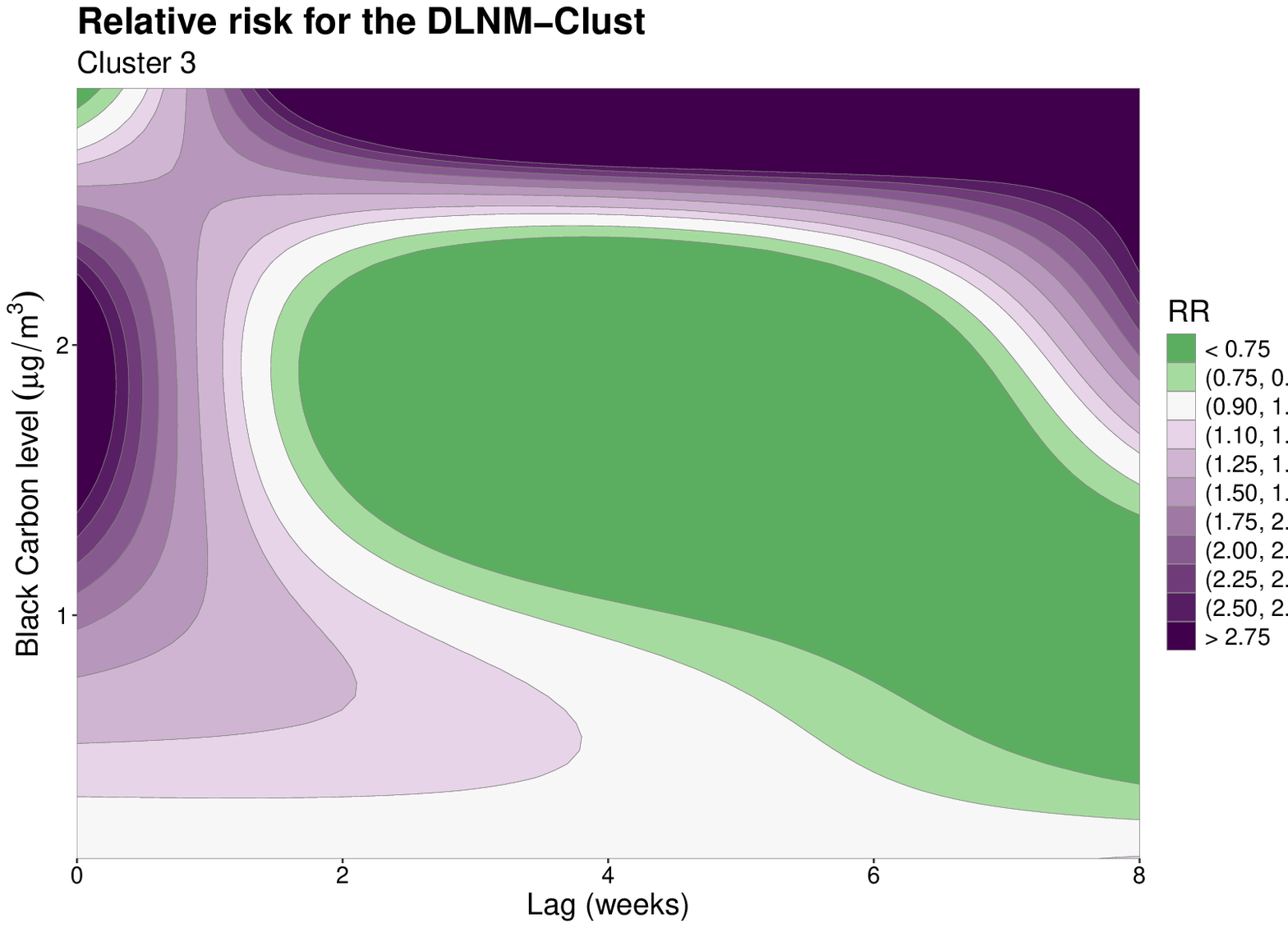}}
\subfloat[]{\includegraphics[width=5.2cm,angle=-90]{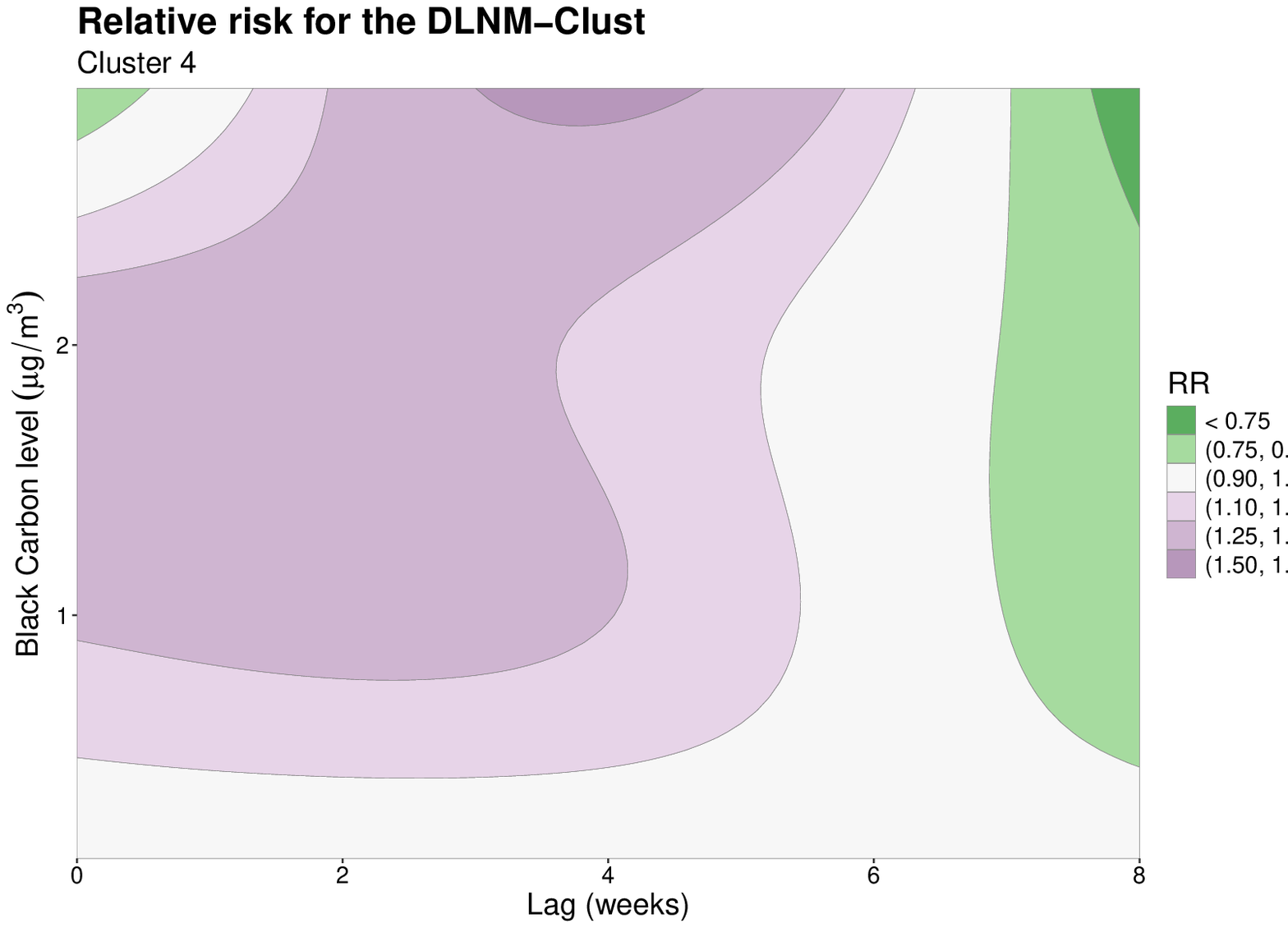}}\\
\subfloat[]{\includegraphics[width=5.2cm,angle=-90]{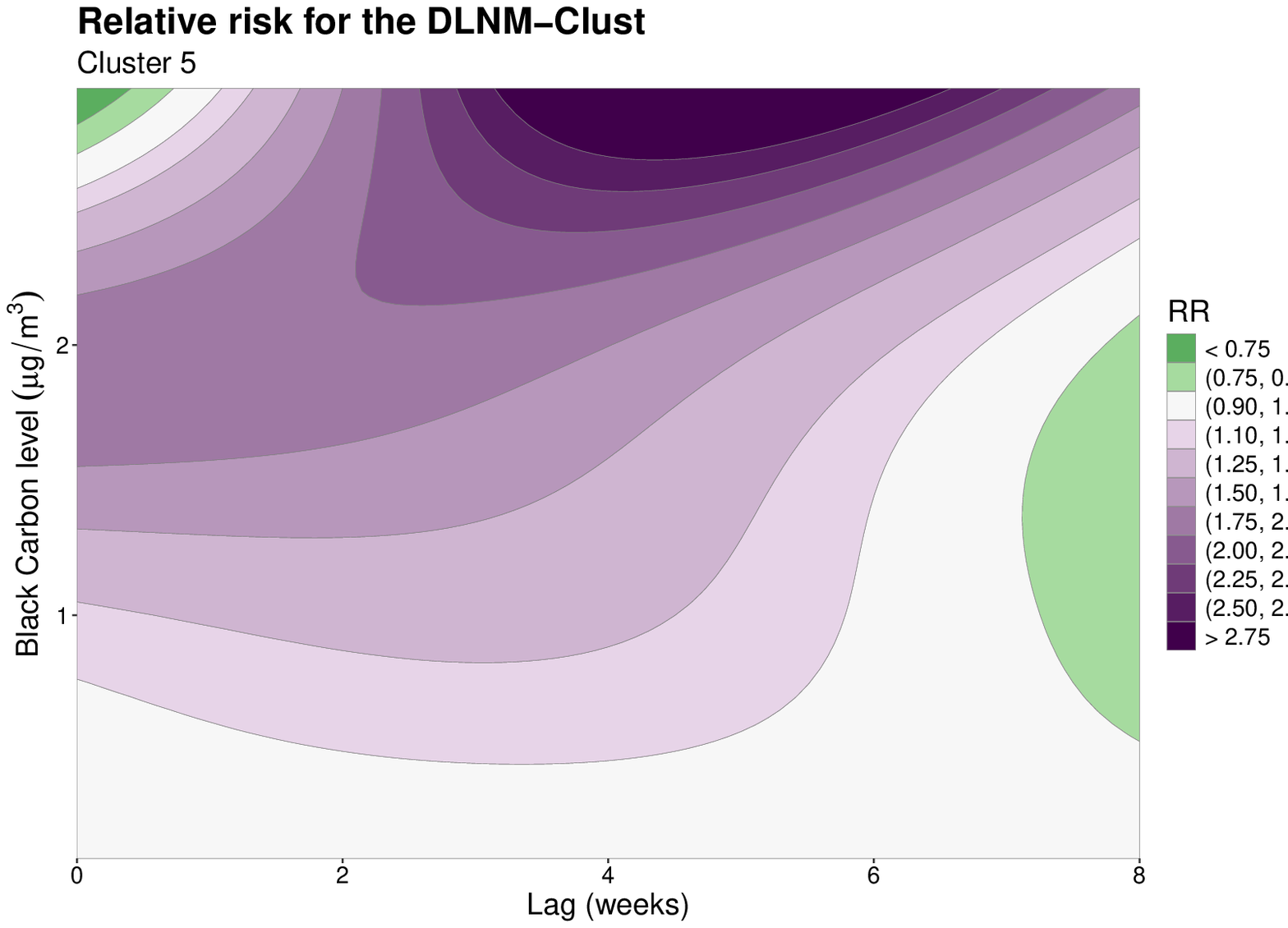}}
\caption{Cluster-specific contour plots of the relative risk of COVID-19 incidence as a function of BC levels at different lags (in weeks) estimated with the DLNM-Clust model for $C=5$ (with spatial assignment). Here we are considering the 5\% sample percentile as the reference value for the BC level}
\label{fig:contours}
\end{figure}

We now turn to showing and analyzing the different exposure-lag-response associations estimated for each of the clusters in the DLNM-Clust model. Figure \ref{fig:contours} displays the contour plots of the relative risk of COVID-19 incidence estimated for the DLNM-Clust model with $C=5$ and spatial clustering assignment, using the 5\% sample percentile of BC levels as the reference value. Five distinguishable exposure-lag-response associations can be observed.

In particular, Clusters 1 and 4 exhibit a relatively planar relative risk surface, showing some resemblance to the overall exposure-lag-response association estimated by the standard DLNM model. In contrast, Clusters 2, 3, and 5 correspond to stronger exposure-lag-response associations. The pattern for Cluster 5 is similar to that inferred by the DLNM model, although the relative risk estimates are higher in this cluster. Differences with respect to the DLNM model are more pronounced for Clusters 2 and 3. For Cluster 2, the highest relative risk values concentrate at the upper BC levels with a lag of 2–4 weeks, whereas in Cluster 3, elevated relative risk estimates extend across the full 8-week lag period considered in the model.

\begin{figure}
    \centering
    \includegraphics[width=0.5\linewidth,angle=-90]{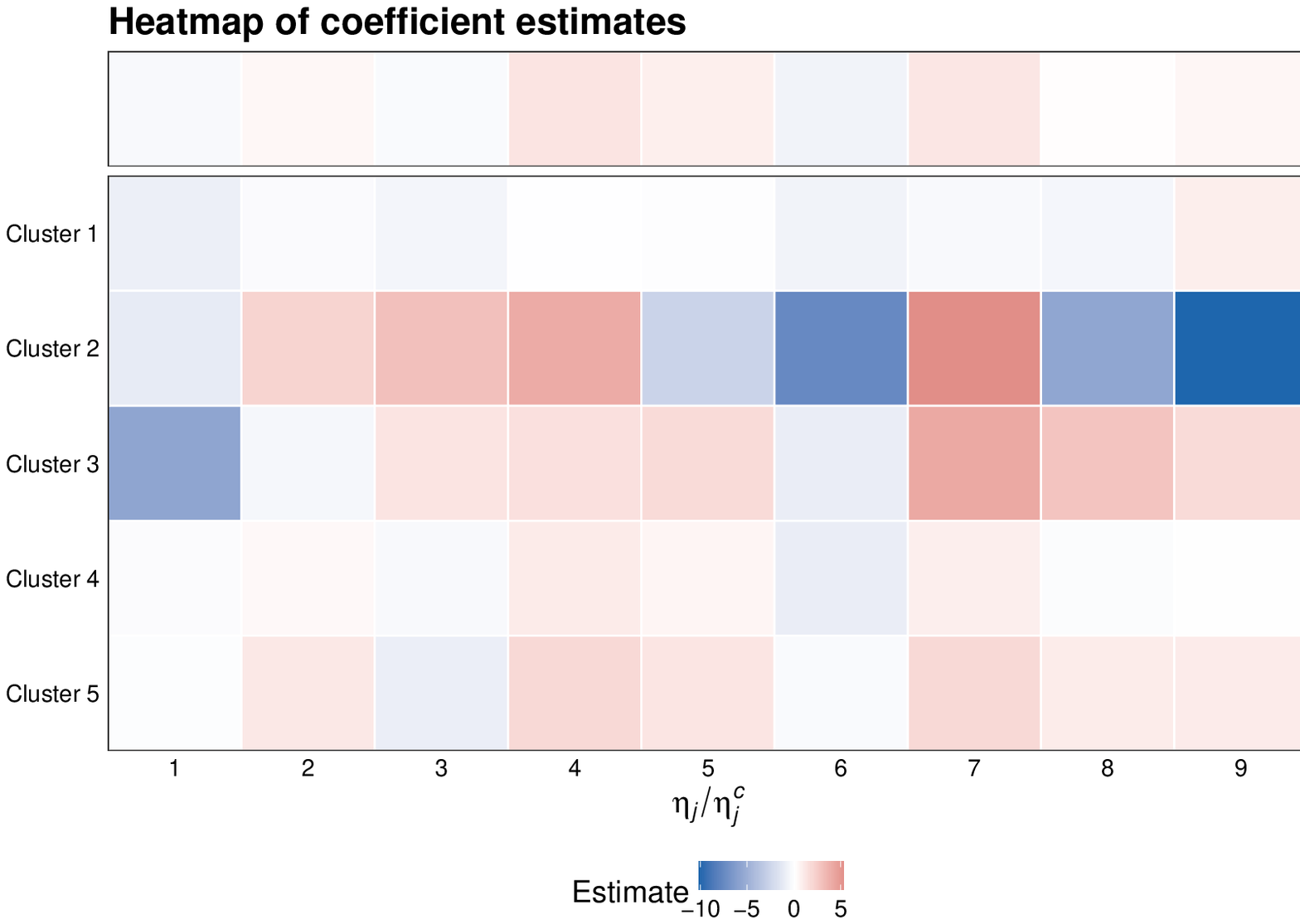}
    \caption{Heatmap of coefficient estimates describing the exposure-lag-response relationship in the standard DLNM and in the specific cluster of the DLNM-Clust model for $C=5$ (with spatial assignment). Estimates correspond to the posterior means of the respective elements of $\boldsymbol{\eta}$ or $\boldsymbol{\eta}_c$, accordingly}
    \label{fig:heatmap}
\end{figure}

Although these results correspond to relative risk estimates based on a single reference value (the 5\% sample percentile of BC levels), similar surface shapes are expected for other reference values. As an additional comparison, Figure \ref{fig:heatmap} displays the estimates of the $\boldsymbol{\eta}$ and $\boldsymbol{\eta}_c$ coefficients obtained for the DLNM and DLNM-Clust models, respectively. Hence, Figure \ref{fig:heatmap} complements Figure \ref{fig:contours} by providing a visual summary of the similarity across clusters through the estimated cross-basis coefficients. This visualization serves to justify the clustering structure by allowing for a direct assessment of the level of separation between clusters. By evaluating these coefficients, we can infer that Cluster 1 captures the mildest exposure-lag-response association. Moreover, while the exposure-lag-response surface of Cluster 4 aligns closely with the standard DLNM estimate, Clusters 2 and 3 exhibit the most pronounced departures from this baseline.

\begin{figure}
    \centering
    \includegraphics[width=0.5\linewidth,angle=-90]{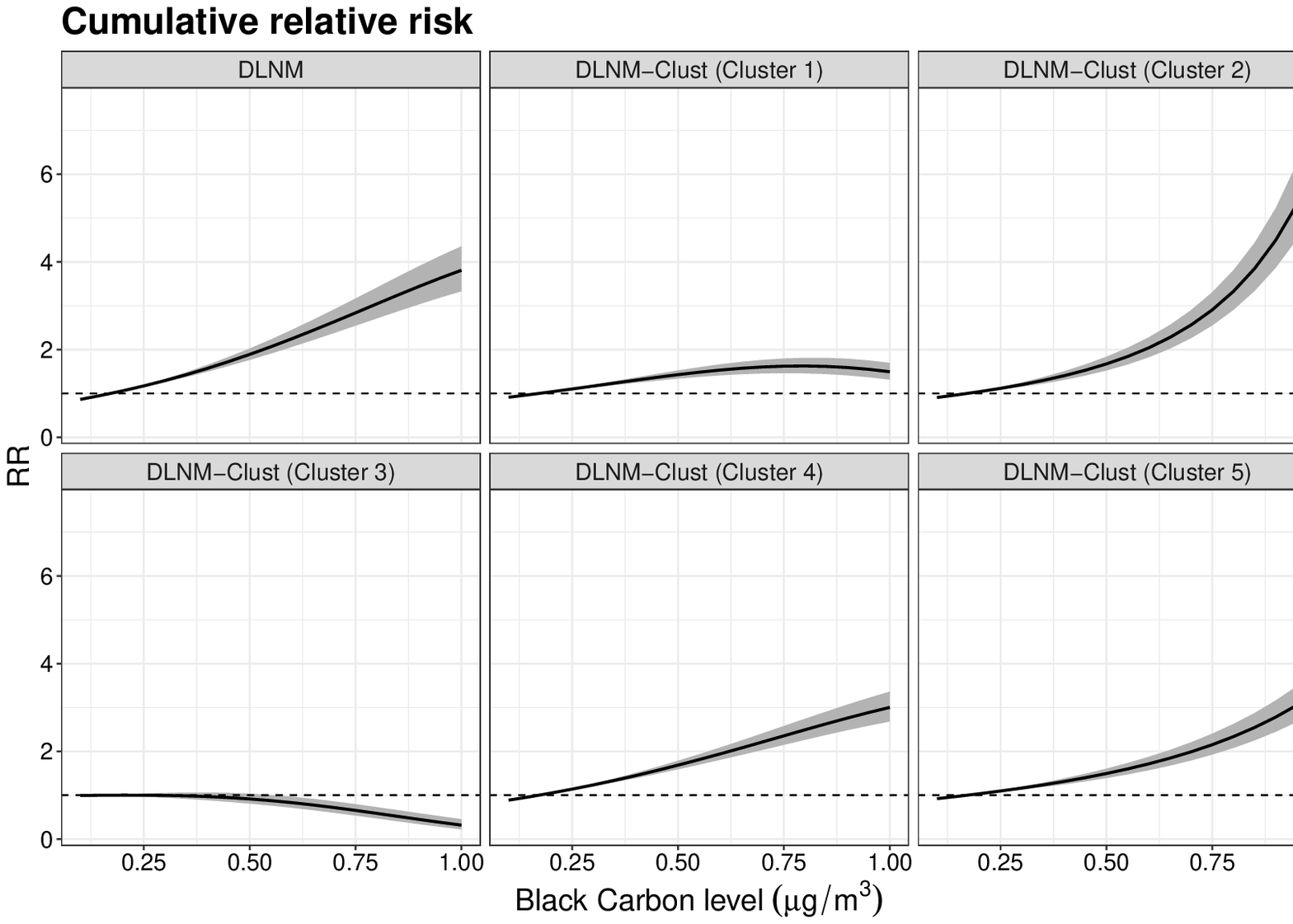}
    \caption{Cumulative relative risk estimates over the 8 weeks of temporal lag considered, for values of the BC level up to 1 $\mu \text{g}/\text{m}^3$, considering the DLNM and the DLNM-Clust for $C=5$ (with spatial assignment). The 95\% confidence band associated with the estimates is also provided}
    \label{fig:cumeffects}
\end{figure}

Finally, we present the cumulative relative risk over the 8-week temporal lag considered in the study, both for the standard DLNM and for each cluster obtained from the DLNM-Clust model with $C=5$. Figure \ref{fig:cumeffects} shows the estimated cumulative relative risk in each case, along with the corresponding 95\% confidence intervals. We focus on cumulative risk up to BC levels of 1 $\mu \text{g}/\text{m}^3$, as this value corresponds approximately to the 90th percentile of observed pollutant levels during the study period.

The results indicate that the cumulative risk for Cluster 2 is substantially higher than that estimated by the DLNM for BC levels near 1 $\mu \text{g}/\text{m}^3$. Conversely, Clusters 1 and 3 exhibit considerably lower cumulative risks than the DLNM estimates. In particular, Cluster 3 shows a protective effect (cumulative risk below 1) for BC levels above 0.5 $\mu \text{g}/\text{m}^3$. This finding may reflect the influence of confounding variables that correlate with these pollutant levels in the municipalities assigned to Cluster 3. Among other factors, it is possible that other atmospheric or environmental variables are interfering, or that municipalities associated with cluster 3 adopted other preventive measures than those in other areas of the country.

\section{Discussion and conclusions}
\label{discussion}

In this article, we have proposed and evaluated a mixture model of DLNMs, which we have called DLNM-Clust. Specifically, the model has been designed to be applied to datasets that include multiple spatially-indexed time series, where it makes sense to consider whether the variation in spatial location implies a different exposure-lag-response relationship. While the present study considers COVID-19 data from Belgium, the underlying framework is designed to be domain-agnostic and broadly generalizable. The DLNM-Clust model serves as a versatile tool for any scenario characterized by non-linear, delayed exposure–response relationships that exhibit spatial heterogeneity. Such dynamics are frequently encountered across various fields, ranging from environmental epidemiology to the social sciences. To facilitate broader adoption and cross-disciplinary application, we have provided the NIMBLE codes implemented in R, allowing practitioners to adapt the model to diverse geographic scales and various health outcomes.

Following the usual principle of Tobler's first law of geography, which states that ``everything is related to everything else, but near things are more related than distant things'' \citep{tobler1970computer}, we have considered two versions of DLNM-Clust. The first, and more basic, version allows each area of the study window to be assigned to any of the clusters, without taking into account the cluster assignment given to neighboring areas. The second version, however, induces neighboring areas to be assigned to the same cluster. This results in more coherent and tightly connected clusters over the study area, facilitating the interpretation of the results. In addition to this, our current framework could be extended to incorporate spatial covariates, such as socioeconomic or demographic indicators. By adopting a ‘spatio-environmental’ autocorrelation perspective \citep{ejigu2020introducing}, the model could cluster areas with similar characteristics, such as income or age distributions, even if they are not geographically contiguous. This direction mirrors recent developments in the field, such as the covariate-dependent DLNM structures explored by \cite{mork2024heterogeneous}.

Furthermore, in the case study carried out, the version of DLNM-Clust with spatial dependence in the assignment of areas to clusters has shown better performance, both from the perspective of the compromise between fit and complexity, and also in terms of the certainty of the assignment, for which the concept of entropy has been used. In addition to this, the results highlighted significant spatial heterogeneity in the exposure-lag-response association, with distinct patterns emerging across different clusters. This heterogeneity was reflected in the varying shapes and magnitudes of the exposure-lag-response curves, underscoring the importance of accounting for spatial variability in the analysis. 

The uncertainty of cluster assignment was also evaluated in the case study. Under a Bayesian framework, the posterior probability of each region belonging to each cluster was calculated based on the MCMC output. These probabilities provide a measure of the uncertainty associated with the cluster assignment for each region. The results showed that some areas had high posterior probabilities of belonging to a particular cluster, indicating a strong association with that cluster. In contrast, other areas had more ambiguous cluster assignments, with posterior probabilities spread across multiple clusters. 

Thus, by examining these posterior probabilities, researchers can gain a deeper understanding of the relationships between the areal units and the clusters, and identify outliers or specific areas with ambiguous cluster assignments. This information can be used to refine the model, explore alternative cluster structures, or investigate the underlying factors driving the observed patterns. At the same time, it is worth noting that the MCMC procedure is also susceptible to the ``label switching'' phenomenon \citep{stephens2000dealing}, where cluster labels may permute during the sampling process. We monitored this by inspecting the trace plots of the $z_i$ latent parameters, finding no evidence of such transitions. However, we acknowledge that label switching remains a potential challenge in the estimation of this kind of mixture-based model.

Another issue to consider is the computational cost of the proposed model. Although the implementation using MCMC is generally expensive, it is not necessarily prohibitive. Exploring more efficient implementations is a potential area for future research. One possibility is to implement this model using the Integrated Nested Laplace Approximation (INLA) proposed by \cite{rue2009approximate}, even though, given the nature of the mixture model, it would be necessary to combine INLA with MCMC \citep{gomez2018markov}.

Finally, we should also acknowledge some limitations of the proposed model. Firstly, the DLNM-Clust approach assumes that there are $C$ distinct exposure-lag-response associations within the spatial study window. However, this may not be a reasonable assumption in certain environmental or epidemiological scenarios where a high level of heterogeneity is unrealistic. For this reason, defining a new DLNM framework based on a single main exposure-lag-response association, from which some areas may deviate according to a clustered random effect, seems to be a promising alternative. Secondly, choosing the value of $C$ is crucial and challenging. In our case study, we took the standard approach of fitting the model to multiple values of $C$ to evaluate the impact of this selection on the performance of the model. A more sophisticated option would be to assign a prior distribution to $C$ and estimate the optimal number of clusters simultaneously. However, this would require well-motivated priors from an epidemiological point of view, which is generally not simple. Thirdly, estimating exposure-lag-response associations for spatially-indexed data in combination with spatial random effects is not immune to spatial confounding \citep{hodges2010adding}. Adding some recently developed techniques for mitigating spatial confounding to the DLNM framework also appears highly convenient.

In conclusion, the results of this study demonstrate the value of the proposed DLNM-Clust model to capture the complex spatially-varying relationships between environmental exposure and health outcomes. The employment of this model could then have implications for the development of more accurate and informative strategies for dealing with environmental health hazards.

\section*{\textbf{Supplementary material}}
The R codes to reproduce the main results of the paper can be found in the GitHub repository \href{https://github.com/albrizre/DLNM-Clust}{https://github.com/albrizre/DLNM-Clust}. In particular, this repository contains the NIMBLE codes corresponding to the different versions of the proposed DLNM-Clust model.

\section*{Declaration of generative AI in scientific writing}
During the preparation of this work the authors used Google Gemini in order to improve the readability of some parts of the paper. After using this tool/service, the authors reviewed and edited the content as needed and take full responsibility for the content of the 
publication.

\bibliography{Bibliography}

\end{document}